\documentclass[prl,twocolumn]{revtex4}

\usepackage{amsmath}
\usepackage{graphicx}
\usepackage{times}

\begin{document}

\title{Topologically-Protected Qubits from a Possible Non-Abelian
Fractional Quantum Hall State}

\author{Sankar Das Sarma$^{1}$, Michael Freedman$^2$, Chetan Nayak$^{2,3}$}
\affiliation{$^1$ Department of Physics, University of Maryland, College Park, MD 20742\\
$^2$Microsoft Research, One Microsoft Way,
Redmond, WA 98052\\
$^3$ Department of Physics and Astronomy, University of California,
Los Angeles, CA 90095-1547}

\date{\today }

\begin{abstract}
The Pfaffian state is an attractive candidate for the observed quantized Hall
plateau at Landau level filling fraction $\nu=5/2$. This is particularly intriguing because this state
has unusual topological properties, including quasiparticle
excitations with non-Abelian braiding statistics.
In order to determine the nature of the $\nu=5/2$ state,
one must measure the quasiparticle braiding statistics. Here, we propose an
experiment which can simultaneously determine the
braiding statistics of quasiparticle excitations and, if they prove to be non-Abelian,
produce a topologically-protected qubit on which a logical NOT operation
is performed by quasiparticle braiding. Using the measured excitation gap
at $\nu=5/2$, we estimate the error rate to be $10^{-30}$ or lower.
\end{abstract}

\maketitle

\paragraph{Introduction}
The computational power of a quantum-mechanical Hilbert space
is potentially far greater than that of any classical device \cite{Shor94,Nielsen00}.
However, it is difficult to harness it because much of the quantum information
contained in a system is encoded in phase relations which
one might expect to be easily destroyed by its interactions with the outside world
(`decoherence' or `error').
Therefore, error-correction is particularly important for quantum computation.
Fortunately, it is possible to represent information redundantly
so that errors can be diagnosed and corrected \cite{Shor95,Gottesman97}.

An interesting analogy with topology suggests itself:
local geometry is a redundant way of encoding topology.
Slightly denting or stretching a surface such as a torus does not
change its genus, and small punctures can be easily repaired
to keep the topology unchanged. Only large changes in the local
geometry change the topology of the surface. Remarkably, there are states
of matter for which this is more than just an analogy. A system
with many microscopic degrees of freedom can have ground
states whose degeneracy is determined by the topology of
the system. The excitations of such a system have exotic braiding
statistics, which is a topological effective interaction between them \cite{stat}.
Such a system is said to be in a topological phase \cite{Wen90a,Wen90b}. The unusual
characteristics of quasiparticles in such states can lead to remarkable
physical properties, such as a fractional quantized Hall
conductance \cite{DasSarma97}.
Such states also have intrinsic fault-tolerance \cite{Kitaev97}.
Since the ground states are sensitive only to the topology of the system,
interactions with the environment, which are presumably local, cannot
cause transitions between ground states unless the environment supplies
enough energy to create excitations which can migrate across the system
and affect its topology. When the temperature is low compared to the
energy gap of the system, such events will be exponentially rare. 

A different problem now arises: if the quantum information is so
well-protected from the outside world, then how can we -- presumably
part of the outside world -- manipulate it to perform a computation?
The answer is that we must manipulate the topology of the system.
In this regard, an important distinction must be made between different
types of topological phases. In the case of those states which are Abelian,
we can only alter the phase of the state by braiding quasiparticles.
In the non-Abelian case, however, there will be a set of $g>1$ degenerate states,
$\psi_a$, $a=1,2,\ldots,g$ of particles at ${x_1}, {x_2}, \ldots, {x_n}$.
Exchanging particles 1 and 2 might do more than just change
the phase of the wavefunction. It might rotate it into a different
one in the space spanned by the $\psi_a$s:
\begin{equation}
{\psi_a} \rightarrow {M^{12}_{ab}}\,{\psi_b}
\end{equation}
On the other hand, exchanging particles 2 and 3
leads to ${\psi_a} \rightarrow {M^{23}_{ab}}\,{\psi_b}$.
If ${M^{12}_{ab}}$ and ${M^{23}_{ab}}$ do
not commute (for at least some pairs of particles),
then the particles obey {\it non-Abelian braiding statistics}.
In the case of a large class of states, the repeated application
of braiding transformations ${M^{ij}_{ab}}$
allows one to approximate any desired
unitary transformation to arbitrary accuracy and, in this sense, they
are universal quantum computers \cite{Freedman02a}.
Unfortunately, no non-Abelian topological states have been
unambiguously identified so far. Some proposals have been put
forward for how such states might arise in highly frustrated magnets
\cite{Freedman03b,Levin04}, where such states might be stabilized by
very large energy gaps on the order of magnetic exchange couplings,
but the best prospects in the short run are in quantum Hall systems,
where Abelian topological phases are already known to exist.
The best candidate is the quantized Hall plateau with
$\sigma_{xy}=\frac{5}{2}\,\frac{e^2}{h}$.
The $5/2$ fractional quantum Hall state (as well as its particle-hole
symmetric analog, the $7/2$ state) is now routinely observed \cite{Xia04} in
high-quality (i.e. low-disorder) samples.  In addition, extensive numerical
work \cite{Morf98} using finite-size diagonalization and wavefunction overlap
calculations indicates that the $5/2$ state belongs to the non-Abelian topological
phase characterized by a Pfaffian quantum Hall
wavefunction \cite{Moore91,Greiter92}.
The set of transformations generated by braiding quasiparticle
excitations in the Pfaffian state is not computationally universal
(i.e. is not dense in the unitary group), but other non-Abelian
states in the same family are.
Thus, it is important to (a) determine if the $\nu=5/2$ state
is, indeed, in the Pfaffian universality class and, if so, to
(b) use it to store and manipulate quantum information.
In this paper, we propose an experimental device which
can address both of these. Features of our device are inspired by anti-dot experiments
measuring the charge of quasiparticles \cite{Goldman95} in Abelian fractional
quantum Hall states such as $\nu=1/3$ and proposals for measuring their statistics \cite{Chamon97}.
Our measurement procedure relies upon quantum interference
as in the electronic Mach-Zehnder interferometer 
in which Aharonov-Bohm oscillations were observed in
a two-dimensional electron gas \cite{Ji03}.

In order to establish which topological phase the $\nu=5/2$ plateau
is in, one must directly measure quasiparticle braiding statistics.
Remarkably, this has never been done even in the case of the usual $\nu=1/3$
quantum Hall plateau (although in this case, unlike in the $\nu=5/2$ case, computational
solutions of small systems leave little doubt about which topological phase the
plateau is in). Part of the problem is that it is difficult to disentangle
the phase associated with braiding from the phase which charged particles
accumulate in a magnetic field \cite{Chamon97}. Ironically, it may actually be easier
to measure the effect of non-Abelian braiding statistics because it is not just a phase
and is therefore qualitatively different from the effect of the magnetic field.

\paragraph{Pfaffian Facts}
To make this latter point clear, let us summarize some important
properties of quasiparticles in the Pfaffian state. The Pfaffian state
may be viewed as a quantum Hall state of $p$-wave paired fermions.
The quasiparticles in this phase have charge-$e/4$ (not $e/2$, as
one might naively assume from the Landau-level filling fraction
$\nu=2+\frac{1}{2}$; this emphasizes the importance of an experiment
such as \cite{Goldman95} to measure the quasiparticle charge at
$\nu=5/2$). When there are $2n$ quasiparticles at fixed positions
in the system, there is a $2^{n-1}$-dimensional degenerate space of states.
Exchanging and braiding quasiparticles is related to the action of the
$2n$-dimensional Clifford algebra on this space \cite{Nayak96c}, as has
recently been confirmed by direct numerical evaluation of the Berry matrices \cite{Tserkovnyak03}.
In particular, two charge-$e/4$ quasiparticles can `fuse' to form a charge-$e/2$ quasiparticle
either with or without a neutral fermion in its core.
One may view the charge-$e/2$ quasiparticle as the quantum Hall
incarnation of a superconducting vortex with a fermionic
zero mode in its core \cite{Read96,Read00,Ivanov01,Stern04}. We will regard the presence or
absence of a neutral fermion in this core state if the two charge-$e/4$
quasiparticles were fused as our qubit. So long as the two
quasiparticles are kept far apart, the neutral fermion is not localized
anywhere and, therefore, the qubit is unmeasurable by any local probe or
environment. However, we can measure the qubit
by encircling it with a charge-$e/4$ quasiparticle.
{\it The presence of the neutral fermion
causes the state to acquire an extra factor of $-1$ during this process.}
The qubit can also be manipulated by taking another charge-$e/4$ quasiparticle
between the two charge-$e/4$ quasiparticles comprising the qubit, i.e. around one
but not the other. Such a process transforms a state without a neutral fermion
into a state with one and vice versa. Thus, it flips the qubit
(and also multiplies by $i$). By performing an
experiment which measures this qubit, flips it, and then re-measures it, we
can demonstrate that the $\nu=5/2$ state is in a non-Abelian topological phase.
(A few additional similar experiments would be necessary to fully nail down
that it is in the Pfaffian phase rather than another non-Abelian phase.)
Such an experiment can only work if the environment does not flip
the qubit before we have a chance to measure it, so the success of this
experiment would demonstrate the stability of a topological qubit
in a non-Abelian quantum Hall state. By varying the time between measurements,
one could determine the decoherence time of the qubit in order to quantitatively
compare it with other approaches to quantum computation.

The claimed quasiparticle braiding properties can be seen from the form of the four-quasihole
wavefunctions given in \cite{Nayak96c}. The ground state wavefunction takes the
form \cite{Moore91,Greiter92}
\begin{equation}
\Psi_{\rm g.s.} (z_j) ~=~ \prod_{j<k} (z_j - z_k)^2 \prod_j e^{- |z_j|^2/4 }
  \cdot {\rm Pf~}\left( {1\over z_j - z_k }\right)~.
\label{grdstate}
\end{equation}
where the Pfaffian is the square root of the determinant of
an antisymmetric matrix. If we write
\begin{multline}
{\Psi_{(13)(24)}}(z_j)= \prod_{j<k} (z_j - z_k)^2 \prod_j e^{- |z_j|^2/4 }
\,\times\\
{\rm  Pf~  }\left( { (z_j - {\eta_1}) (z_j - {\eta_3} )(z_k
-{\eta_2} ) (z_k - {\eta_4} ) + (j \leftrightarrow k )
\over z_j - z_k}\right)
\label{qhwf}
\end{multline}
and similarly for ${\Psi_{(14)(23)}}$, then the four-quasihole
wavefunctions can be written in a basis in which their braiding
is completely explicit:
\begin{multline}
\label{eqn:fourqh}
{\Psi^{(0,1)}}(z_j) =  { {\left({\eta_{13}}{\eta_{24}}\right)^{1\over 4}}
\over{(1 \pm \sqrt{x})^{1/2}}}\,
\left( {\Psi_{(13)(24)}} \,\,\pm\,\,\sqrt{x}\,\,
{\Psi_{(14)(23)}}\right)
\end{multline}
where $\eta_{13}={\eta_1}-{\eta_3}$, etc. and
$x~=~\eta_{14}\eta_{23}/\eta_{13}\eta_{24}$.
Let us suppose that the quasiholes at $\eta_1$ and $\eta_2$ form our
qubit. The quasiholes at $\eta_3$ and $\eta_4$ will be used to measure
and manipulate them. From (\ref{eqn:fourqh}), we see that taking $\eta_3$ around
$\eta_1$ and $\eta_2$ results in a factor $i$ in the state ${\Psi^{(0)}}$
but $-i$ in the state ${\Psi^{(1)}}$. Taking $\eta_3$ around either
$\eta_1$ or $\eta_2$ (but not both) transforms ${\Psi^{(0)}}$ into
$i\,{\Psi^{(1)}}$ and vice versa.

\begin{figure}[tbh]
\includegraphics[width=3.45in]{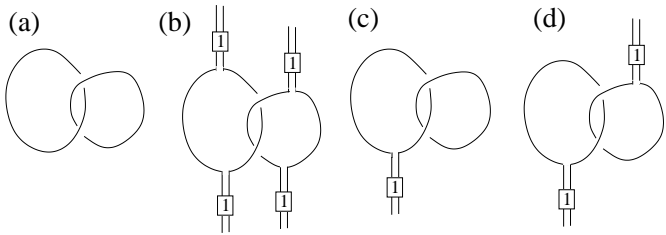}
\caption{By evaluating the Jones polynomial at $q=\exp(\pi i/4)$ for these links,
we can obtain the desired matrix elements for braiding operations manipulating the qubit.
The boxed $1$ is a projector on the pair of quasiparticles which puts them in the
state $|1\rangle$.}
\label{fig:Jones-qubit}
\end{figure}

It is also possible \cite{Fradkin98} to verify the logic associated to braiding operations
using a few formal properties of the Jones polynomial at $q=\exp(\pi i/4)$.
Taking one quasiparticle around the qubit pair (`linking') results
in an extra $-1$ if the qubit is in state $|1\rangle$
(a factor $d=-q-q^{-1}$ also arises regardless of whether or not the quasiparticle
encircles the qubit). The Jones polynomial (operator) at $q=\exp(\pi i/4)$ vanishes
for the links in figures \ref{fig:Jones-qubit}a,b by calculation, \ref{fig:Jones-qubit}c
by parity, and is non-vanishing only for \ref{fig:Jones-qubit}d
(which applies to all processes with topologically-equivalent link diagrams,
e.g. interchanging inputs/outputs so, for example, \ref{fig:Jones-qubit}d
corresponds to four different processes).
In case \ref{fig:Jones-qubit}d, the qubit is flipped by the elementary braid operation.

\begin{figure}[tbh]
\includegraphics[width=3.45in]{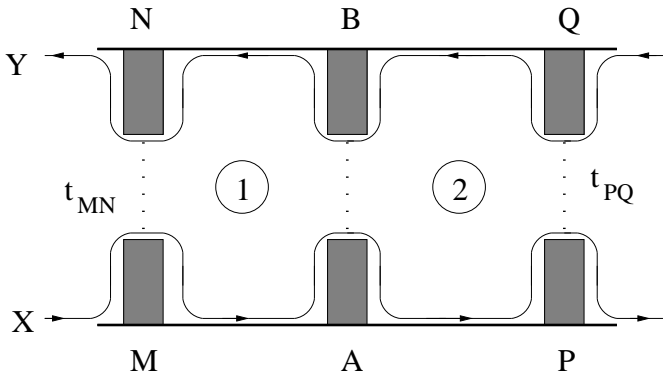}
\caption{A schematic depiction of a Hall bar with front gates which enable
tunneling between the two edges at M, N and P, Q, thereby allowing a
measurement of the qubit formed by the correlation between anti-dots 1 and 2.
Front gates (shaded regions) also allow tunneling at A, B which flips the qubit.}
\label{fig:pfaff-dev}
\end{figure}

\paragraph{Experimental Configuration}
The basic setup which we propose is a quantum Hall bar with two
individually-gated anti-dots in its interior, labeled 1 and 2 in
figure \ref{fig:pfaff-dev}. There are front gates which enable tunneling
between A and B at the edges. It is useful to have a third anti-dot at the point C
midway between A and B in order to precisely control the charge which tunnels
between A and B, but we have not depicted it to avoid clutter.
Two more front gates enable tunneling at M and N and at P and Q.
There are three basic procedures which we would like to execute:
(1) initialize the qubit and measure its initial state, (2) flip the qubit,
and (3) measure it again.

In order to initialize the qubit, we first put charge $e/2$ on one of the antidots,
say 1. Since the fermionic zero mode is now localized on this antidot, the environment
will `measure' it, and it will either be occupied or unoccupied (not a superposition
of the two). We can determine which state it is in by applying voltage to
the front gates at M and N and at P and Q so that tunneling can occur there with
amplitudes $t_{MN}$ and $t_{PQ}$. The longitudinal conductivity, $\sigma_{xx}$
is determined by the probability for current entering the bottom edge at X
in figure \ref{fig:pfaff-dev} to exit along the top edge at Y. This is given,
to lowest order in $t_{MN}$ and $t_{PQ}$, by the interference between two
processes: one in which a quasiparticle tunnels from M to N;
and another in which the quasiparticle instead continues along the bottom
edge to P, tunnels to Q, and then moves along the top edge to N.
(We subsume into $t_{PQ}$ the phase associated with the extra distance
travelled in the second process.) The relative phase of these processes
depends on the state of the qubit. If a neutral fermion is not present, which we
will denote by $|0\rangle$, then $\sigma_{xx}\propto |t_{MN}+i\,t_{PQ}|^2$.
If it is present, however, which we denote by $|1\rangle$, then
$\sigma_{xx}\propto |t_{MN}-i\,t_{PQ}|^2$. We take the visibility of Aharonov-Bohm
oscillations in a device with similar limitations \cite{Ji03} (e.g. the possibility of
the tunneling quasiparticles becoming dephased
by their interaction with localized two-level systems)
as an indication that our proposed read-out procedure will work.

Without loss of generality, let us suppose that the initial state of the qubit
is $|0\rangle$. Now, let us apply voltage to anti-dots 1 and 2 so that
charge $e/4$ is transferred from 1 to 2. There is now one charge-$e/4$ quasihole
on each anti-dot. The state of the qubit is unaffected by this process.
In order to flip this qubit, we now apply voltage to the
front gates at A and B so that one charge $e/4$ quasiparticle tunnels
between the edges. In order to ensure that only a single quasiparticle tunnels,
it is useful to tune the voltage of the anti-dot at C and the backgate at A so that
a single quasiparticle tunnels from the edge to the anti-dot at C.
(If the anti-dot is small, its charging energy will be too high to allow more than
one quasiparticle to tunnel at once.) We can then lower the voltage of the backgate
at A so that no further tunneling can occur there and apply voltage to the backgate
at B so that the quasiparticle can tunnel from C to B. By this two-step process,
we can tunnel a single quasiparticle from A to B. If the $\nu=5/2$ plateau
is in the phase of the Pfaffian state, this will transform $|0\rangle$ to $|1\rangle$.
This is our logical NOT operation.
The gate which creates the anti-dot at C must be turned off
at the beginning and end of the bit flip process so that there are no quasiparticles
there either before or after which could become entangled with our qubit.

We can now measure our qubit again by tuning the front gates so that
tunneling again occurs between M and N and between P and Q with amplitudes
$t_{MN}$ and $t_{PQ}$. If, as expected, the qubit is now in the state $|1\rangle$
we will find $\sigma_{xx}\propto |t_{MN}-i\,t_{PQ}|^2$. On the other hand, if
the $\nu=5/2$ state were Abelian, $\sigma_{xx}$ would not be affected by the
motion of a quasiparticle from A to B.

In order to execute these steps, it is important that we know that
we have one (modulo 4) charge-$e/4$ quasihole on each anti-dot. This can be ensured
by measuring the tunneling conductance $G_t^{\rm ad}$ from one edge to the other
through each anti-dot \cite{Goldman95}. As we sweep the magnetic field,
there will be a series of peaks in $G_t^{\rm ad}$ corresponding to the passage
through the Fermi level of quasihole states of the antidot.
The spacing $\Delta\! B$ between states is determined by the condition that
an additional state passes through the Fermi level when one additional half-flux-quantum,
${\Phi_0/}2$ is enclosed in the dot. Thus, the number of quasiholes is given
simply by $\lfloor B/\Delta\! B\rfloor$. Alternatively, with a back gate, we could directly
measure capacitatively the charge on each anti-dot \cite{Goldman95}.
If the back gate voltage is $V_{BG}$ (relative to the zero quasihole
case when the gate defining the anti-dot is turned off), then the charge on
the anti-dot is $q = \epsilon V_{BG} A/d$, where $A={\Phi_0}/2\Delta\! B$
is the area of the dot, $\epsilon$ the dielectric constant, and $d$ the distance
between the back gate and the 2DEG.

\paragraph{Estimate of Error Rate}
Bit flip and phase flip errors, respectively, occur when an uncontrolled
charge-$e/4$ quasiparticle performs one of the two basic processes above: encircling
one of the anti-dots (or passing from one edge to the other between them)
or encircling both of them. The rate for these processes is
related to the longitudinal resistivity (which vanishes within experimental
accuracy) because it is limited by the density and mobility of excited quasiparticles.
Even without considering
the suppression factor associated with the latter (which depends on
the ratio of the diffusion or hopping length, $a$, to the system size, $L$),
we already have a strong upper bound on the error rate following
from the thermally-activated form of the former
(in ${k_B}=1$ units):
\begin{equation}
\frac{\Gamma}{\Delta} \sim \frac{T}{\Delta}\, e^{-\Delta/T} < 10^{-30}
\end{equation}
Here, we have used the best current measured value \cite{Pan-comm} for the
quasiparticle gap $\Delta=500$mK of the $5/2$ state and
the lowest achieved measurement temperature $T=5$mK.
For arbitrary braid-based computation, in a more elaborate device, it is sufficient
if we further have $e^{\Delta/T}>\nu\Delta L^2$, where $\nu$ is the density-of-states.
The effect of residual pinned quasiparticles
can be diagnosed and accounted for in software.
These error rates are substantially lower than the estimated
error rate for any other physical implementations of quantum computation in
any proposed architectures.  Compared to other scalable solid state
architectures, such as localized electron spin qubits \cite{DasSarma04} in
Si or GaAs nanostructures, where the estimated error rate is around $10^{-4}$
even in the best possible circumstances, the errors associated with $\nu=5/2$ quantum Hall
anyons is essentially negligible.  This miniscule error rate arises from the intrinsic
robustness of the topological phase which is fundamentally immune to all
local environmental perturbations.

The ideal error rate for the $5/2$ state may actually be substantially lower
than even this very low currently achievable value of $10^{-30}$.  There is
strong theoretical evidence \cite{Morf03} that the ideal excitation gap ($\sim 2$K) for the
$5/2$ quantum Hall state is much larger than the currently achieved gap value
of $500$mK.  Using an ideal gap of $2$K, we get an
astronomically low error rate of $10^{-100}$.  This expected higher value of $\Delta$
($\sim 2$K) is consistent with the experimental development of the activation gap
measurement \cite{Xia04} of the $5/2$ state.  The early measurements on fairly modest
quality samples (i.e. relatively highly disordered) gave $\Delta\sim 100$mK whereas
recent measurements in extremely high-quality (i.e. low disorder) samples
give $\Delta\sim 300-500$mK \cite{Xia04}.  This implies that the $5/2$ excitation gap is
susceptible to strong suppression by disorder as has recently been
theoretically argued \cite{Morf03}.  Since improvement in sample quality has already
led to a factor of $5$ enhancement in $\Delta$ (from $100$mK to $500$mK), it is not
unreasonable to expect further improvements.

There are, in principle, other sources of error, but we expect them to be of
minor significance.  For example, if two quasiparticles come close to each
other, then their mutual interaction leads to an error (e.g. through the
exchange of a virtual particle).  Such a virtual exchange is, however, a
quantum tunneling process which should be exponentially suppressed.
Therefore keeping the quasiparticles reasonably far from each other should
essentially eliminate this error.

We note that, although we have discussed only the $5/2$ Pfaffian quantized
Hall state throughout this paper, our  considerations and arguments apply
equally well to the experimentally often-observed $7/2$ quantized Hall state
which, being the 'hole' analog of the $5/2$ state by virtue of the
particle-hole symmetry, should have equivalent topological and non-Abelian
properties.  We believe the $5/2$ state to be a better experimental candidate
for topological quantum computation because the measured excitation gap in
the $5/2$ state tends to be much higher than that in
the $7/2$ state.  We should also mention that recently \cite{Xia04} the $12/5$
fractional quantum Hall state has been observed experimentally in the
highest mobility sample at the lowest possible temperatures.  This state,
thought to be a non-Abelian state related to parafermions
\cite{Read99}, is particularly exciting from
the perspective of topological quantum computation because its
braid group representation is dense in the unitary
group \cite{Freedman02a} making this state an ideal candidate for topological quantum
computation.  The measured gap value in the 12/5 state is
currently rather small ($\sim 70$mK), making any experimental effort along the
line of our discussion in this paper premature at this stage. However, we expect that
this is also strongly affected by disorder and that the eventual ideal
gap at $12/5$ will be much larger.

\paragraph{Acknowledgements}
We would like to thank J. Eisenstein, A. Kitaev, C. Marcus, and
W. Pan for discussions. We have been supported by
the ARO under Grant No. W911NF-04-1-0236.
C. N. has also been supported by the NSF under
Grant No. DMR-0411800.

\vskip -0.5cm


\end{document}